\documentclass[letterpaper,english,onecolumn, draftcls, 12pt]{IEEEtran}
\usepackage[T1]{fontenc}
\usepackage[latin9]{inputenc}
\usepackage{babel}

\usepackage{verbatim}
\usepackage{amsthm}
\usepackage{amsmath}
\usepackage{graphicx}
\usepackage{setspace}
\usepackage{amssymb}
\usepackage[unicode=true,
 bookmarks=true,bookmarksnumbered=false,bookmarksopen=false,
 breaklinks=false,pdfborder={0 0 1},backref=false,colorlinks=false,pdfpagemode=FullScreen]
 {hyperref}
\hypersetup{
 pdfauthor={Paolo Castiglione - Osvaldo Simeone}}
\usepackage{breakurl}

\makeatletter

\theoremstyle{plain}
\newtheorem{thm}{Theorem}
\theoremstyle{definition}
\newtheorem{example}[thm]{Example}
\theoremstyle{plain}
\newtheorem{prop}[thm]{Proposition}
\theoremstyle{remark}
\newtheorem{rem}[thm]{Remark}

\makeatother

\begin{document}

\title{Energy Management Policies\\
for Energy-Neutral Source-Channel Coding%
\thanks{This work has been submitted to the IEEE for publication. Copyright
may be transferred without notice, after which this version may no
longer be accessible. Part of this paper has been accepted for publication
at the 9th Int. Symposium on Modeling and Optimization in Mobile,
Ad Hoc, and Wireless Networks (WiOpt 2011), Princeton, New Jersey,
USA on May 2011. This work was supported by the Austria Science Fund
(FWF) through grant NFN SISE (S106). The Telecommunications Research
Center Vienna (FTW) is supported by the Austrian Government and the
City of Vienna within the competence center program COMET. %
}}

\author{P. Castiglione\authorrefmark{1}, O. Simeone\authorrefmark{2}, E.
Erkip\authorrefmark{3}, and T. Zemen\authorrefmark{1}\vspace{-1mm}\authorblockA{\\\authorrefmark{1}Forschungszentrum Telekommunikation Wien, Austria} \vspace{-2mm}\authorblockA{\\\authorrefmark{2}CWCSPR, ECE Dept, NJIT, New Jersey, USA}
\vspace{-2mm}\authorblockA{\\\authorrefmark{3}Dept. of ECE, Polytechnic Inst. of NYU, New York, USA}\vspace{-3.4mm}\vspace{-8.5mm}
}
\maketitle
\begin{abstract}
In cyber-physical systems where sensors measure the temporal evolution
of a given phenomenon of interest and radio communication takes place
over short distances, the energy spent for source acquisition and
compression may be comparable with that used for transmission. Additionally,
in order to avoid limited lifetime issues, sensors may be powered
via energy harvesting and thus collect all the energy they need from
the environment. This work addresses the problem of energy allocation
over source acquisition/compression and transmission for energy-harvesting
sensors. At first, focusing on a single-sensor, energy management
policies are identified that guarantee a maximal average distortion
while at the same time ensuring the stability of the queue connecting
source and channel encoders. It is shown that the identified class
of policies is optimal in the sense that it stabilizes the queue whenever
this is feasible by any other technique that satisfies the same average
distortion constraint. Moreover, this class of policies performs an
independent resource optimization for the source and channel encoders.
Analog transmission techniques as well as suboptimal strategies that
do not use the energy buffer (battery) or use it only for adapting
either source or channel encoder energy allocation are also studied
for performance comparison. The problem of optimizing the desired
trade-off between average distortion and delay is then formulated
and solved via dynamic programming tools. Finally, a system with multiple
sensors is considered and time-division scheduling strategies are
derived that are able to maintain the stability of all data queues
and to meet the average distortion constraints at all sensors whenever
it is feasible.
\end{abstract}

\author{\vspace{-8.5mm}
}

\section{Introduction}

%
{}In the {}``smart world'', wireless sensor networks (WSNs) play a
central role in bridging the real and the digital worlds \cite{Economist}.
WSNs are typically designed under the assumptions that communication
resources are limited by the energy available in the battery and that
the most significant source of energy expenditure is radio transmission.
However, modern cyber-physical systems are expected to operate over
a virtually infinite lifetime. This can only be achieved by overcoming
the limitations of battery-powered sensors and allowing the sensors
to \emph{harvest} the energy needed for their operation from the environment,
e.g., in the form of solar, vibrational or radio energy \cite{Par05,cps}.
The regime of operation in which the system operates in a fully self-powered
fashion is referred to as \textit{energy neutral} \cite{Kansal}\emph{.
}Moreover, when sensors are tasked with acquiring complex measures,
such as long time sequences of given phenomena of interest, and when
transmission takes place over small distances, the energy cost of
running the source acquisition system (sensing, sampling, compression)
may be comparable with that of radio transmission \cite{HE06,LU03}.

\begin{figure}[t]
\centering{\includegraphics[clip,width=18cm]{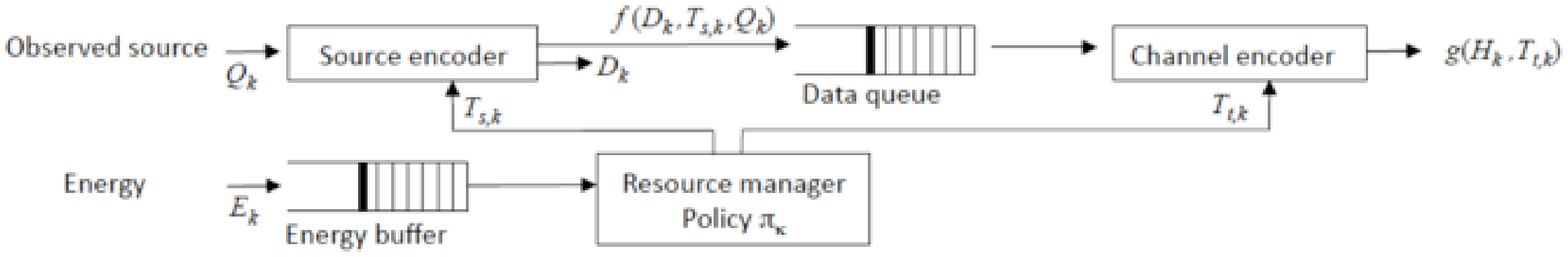}}

\caption{An energy-harvesting sensor composed of a cascade of a source and
a channel encoder powered by a resource manager that allocates the
energy available in the buffer (e.g., battery or capacitor).}
\label{Flo:model}%
\end{figure}

Based on the discussion above, in this paper, we address the problem
of energy management for a WSN in which sensors are powered via energy
harvesting and in which source acquisition and radio transmission
have comparable energy requirements. We first focus on a system with
a single sensor communicating to a single receiver, as shown in Fig.
\ref{Flo:model}, in order to concentrate on the main aspects of the
problem. The sensor is equipped with a battery in which the harvested
energy is stored. In each time slot, the sensor acquires a time sequence
for the phenomenon of interest, which is characterized by a measurement
signal-to-noise ratio (SNR) and autocorrelation, and stores the resulting
bits, after possible compression, into a data queue. At the same time,
it transmits a number of bits from the data queue to the fusion center
over a fading channel with an instantaneous channel SNR. Based on
the statistics of the energy harvesting process, and based on the
current states of the measurement quality, of channel SNR, and of
the data queue, the energy management unit must perform energy allocation
between source acquisition and data transmission so as to optimally
balance competing requirements such as distortion of the reconstruction
at the receiver, queue stability and delay. This optimization problem
is the main subject of this work. We further extend our analysis to
the problem of scheduling multiple sensors that are communicating
to the same receiver.

The model at hand is inspired by the work in \cite{HE06,LU03} and
\cite{Sha10}. In \cite{Sha10}, the energy-harvesting sensor allocates
power to data transmission over different channel SNRs, since the
bit arrival process is assumed to be given and not subject to optimization.
This is unlike our work in which a key problem is that of allocating
resources between transmission and source compression in order to
guarantee given constraints such as distortion and queue stability.
The problem of energy allocation between source compression and transmission
was instead first studied in \cite{HE06,LU03}, but in power-limited
systems with no energy-harvesting capabilities.

Other related works pertain to the study of energy-harvesting WSNs.
This is a growing field with recent significant contributions. Here
we only point to the works that are most related to ours, besides
the ones already mentioned above. An information-theoretic analysis
of a single-sensor system with energy-harvesting is presented in \cite{Oze10,Raj10},
where it is shown that energy-harvesting does not affect the capacity
of the channel, as long as one assumes that the battery has an arbitrarily
large storage capacity. An optimal strategy for a single-sensor system
that can control both the {}``acceptance rate'' of the arriving
bits and the power allocation with the aim of maximizing the throughput
under stability constraints is developed in \cite{Mao10}. Optimal
scheduling is instead studied in \cite{YAN10,Tut10,Sha09}. The effect
of a finite battery is studied in \cite{Sri10}, where the trade-off
between achievable rate and battery discharge probability is characterized.
It is noted that all these works do not model the aspect of source
acquisition and processing.

The main contributions of this work are summarized as follows. (\textit{i})
We propose a simple, but general, model for an energy-harvesting sensor
operating over a time-varying channel (Sec. \ref{sec:System-Model}).
(\textit{ii}) For a single-sensor system, we design a novel class
of \textit{distortion-optimal energy-neutral }resource allocation
policies that are able to stabilize the data queue and, simultaneously,
to meet an average distortion constraint, whenever it is feasible
by any policy (Sec. \ref{sub:Throughput-distortion-optimality}).
For the case where multiple sensors access the same uplink channel
in time division, we identify a distortion-optimal energy-neutral
class of \textit{scheduling} policies (Sec. \ref{sec:Scheduling-Protocols}).
(\textit{iii}) We compare the performance of the optimal policies
with a number of less complex strategies, such as {}``analog'' techniques
\cite{GAS03} (Sec. \ref{sub:Analog-transmission}) and fixed time
division multiple access (TDMA) scheduling strategies (Sec. \ref{sub:Numerical-Results_sched}).
(\textit{iv}) Finally, we formulate the problem of optimizing a desired
trade-off between average delay and distortion, which is solved via
dynamic programming tools (Sec. \ref{sec:Delay-Distortion-Optimization}).

\section{System Model\label{sec:System-Model}}

In this section, we introduce the system model, main assumptions and
problem definition.

We consider a system in which a single sensor communicates with a
single receiver as depicted in Fig. \ref{Flo:model}. The extension
of this system to the case of multiple sensors and a single receiver
is studied in Sec. \ref{sec:Scheduling-Protocols}. In most of the
paper, we assume that the sensor performs separate source and channel
coding, as described in the following. A different approach is considered
in Sec. \ref{sub:Analog-transmission}.

Time is slotted. The energy $E_{k}\in\mathbb{R}_{+}$ harvested in
time-slot \emph{k} is stored in an {}``energy buffer'', also referred
to as battery, with infinite size. For convenience, the energy $E_{k}$
is normalized to the number $N$ of channel discrete-time symbols
available for communication in each time slot, also referred to as
channel uses. The energy arrival $E_{k}$ is assumed to be a stationary
ergodic process. The probability density function (pdf) of $E_{k}$
is $p_{E}\left(e\right)$. The energy $\tilde{E}_{k+1}$ available
for use at slot $k+1$ is the residual energy from the previous slot
plus the energy arrival at time-slot \emph{$k+1$}. This evolves as
\begin{equation}
\tilde{E}_{k+1}=\left[\tilde{E}_{k}-\left(T_{s,k}+T_{t,k}\right)\right]^{+}+E_{k+1},\label{eq:coda1}\end{equation}
where $T_{s,k}$ and $T_{t,k}$ account for the energy spent in slot
$k$ per channel use for source acquisition and data transmission,
respectively, as discussed below. Notice that the energy arriving
at time slot $k+1$ is immediately available for use in that slot.

The sensor measures $M$ samples of a given source during each slot.
The quality of such observation in slot $k$ depends on a parameter
$Q_{k}\in\mathcal{Q}$, which is assumed to be a stationary ergodic
process over the time slots $k$. For instance, the sensor may perform
measurements of the phenomenon of interest whose SNR $Q_{k}$ changes
across blocks $k$ due to source movement or environmental factors
affecting the measurement quality. The set $\mathcal{Q}$ is assumed
to be discrete and finite, and the (stationary) probability mass function
(pmf) for $Q_{k}$ is given by $\Pr(q)=\Pr(Q_{k}=q)$, for $q\in\mathcal{Q}$%
{}. The sensor acquires the source in a lossy fashion. The loss, due
to sampling, analog-to-digital conversion and compression, is characterized
by distortion $D_{k}\in\mathbb{R}^{+}$, as measured with respect
to some distortion metric such as the mean square error (MSE).

The number of bits generated by the \textit{\emph{source encoder}}\emph{
}at the sensor at slot $k$ is $X_{k}=f(D_{k},T_{s,k},Q_{k})$, where
$f$ is a given function of the distortion level $D_{k}$, of the
energy per channel use allocated to the source encoder $T_{s,k}$
and on the observation state $Q_{k}$. The resulting bit stream is
buffered in a first-input-first-output (FIFO) data queue with queue
length $\tilde{X}_{k}$. The function $f(D_{k},T_{s,k},Q_{k})$ is
assumed to be separately continuous convex and non-increasing in $D_{k}$
and $T_{s,k}$. For simplicity, we will denote such functions also
as $f^{q}(D_{k},T_{s,k})=f(D_{k},T_{s,k},Q_{k}=q)$. Some examples
for function $f$ will be provided below in Sec. \ref{sub:Rate-distortion-function}.

The fading channel between sensor and destination is characterized
by a process $H_{k},$ assumed to be stationary ergodic, where $H_{k}\in\mathcal{H}$,
with set $\mathcal{H}$ being discrete and finite in order to ease
the numerical evaluations. We assume a slowly time-variant scenario.
The pmf of $H_{k}$ is given by $\Pr(h)=\Pr(H_{k}=h)$, for $h\in\mathcal{H}$.
The \textit{channel encoder }uses the channel\textit{ }$N$ times
per slot, and the transmission requires $T_{t,k}$ energy per channel
use. A maximum number $g\left(H_{k},T_{t,k}\right)$ of bits per slot
can be delivered successfully to the destination. The channel rate
function $g\left(H_{k},T_{t,k}\right)$ is assumed to be continuous,
concave, non-decreasing in $T_{t,k}$, and $g\left(H_{k},0\right)=0$.
We also use the notation $g^{h}\left(T_{t,k}\right)=g\left(H_{k}=h,T_{t,k}\right)$.
An example is the Shannon capacity on the complex additive white Gaussian
noise (AWGN) channel $g^{h}\left(T_{t,k}\right)=N\times\log(1+hT_{t,k})$
\cite{COV}. We remark that adopting the Shannon capacity implies
the use of rate-adaptive schemes with sufficiently long codewords
so that the block error probability becomes negligible. For the given
function $g(H_{k},T_{t,k})$, the channel encoder takes $\min[\tilde{X}_{k},g\left(H_{k},T_{t,k}\right)]$
bits from the data buffer, using the selected transmission energy
$T_{t,k}$. Note that we do not consider the effects of channel errors
nor the costs of channel encoding/decoding and of channel state information
feedback, which are beyond the scope of the present paper and subject
to future work.

Based on the discussion above, the data queue evolves as\begin{equation}
\tilde{X}_{k+1}=\left[\tilde{X}_{k}-g\left(H_{k},T_{t,k}\right)\right]^{+}+f\left(D_{k},T_{s,k},Q_{k}\right).\label{eq:coda2}\end{equation}
To illustrate the trade-offs involved in the energy allocation between
$T_{t,k}$ and $T_{s,k}$, we remark that, by providing more energy
$T_{s,k}$ to the source encoder, one is able, for the same distortion
level $D_{k}$, to reduce the number $f\left(D_{k},T_{s,k},Q_{k}\right)$
of bits to be stored the data buffer. At the same time, less energy
$T_{t,k}$ is left for transmission, so that the data buffer is emptied
at a lower rate $g\left(H_{k},T_{t,k}\right)$. Viceversa, one could
use less energy to the source encoder, thus producing more bits $f\left(D_{k},T_{s,k},Q_{k}\right)$,
so that more energy would be available to empty the data buffer.

\subsection{Rate-Distortion-Energy Trade-Off\label{sub:Rate-distortion-function}}

In the following, we present some examples for function $f^{q}\left(D_{k},T_{s,k}\right)$,
as available in the literature. Recall that this function provides
the trade-off between the distortion $D_{k}$, the energy consumption
$T_{s,k}$ and the number of bits produced by the source encoder.
\begin{example}
Consider the observation model $R_{k,i}=\sqrt{Q_{k}}U_{k,i}+Z_{k,i}$,
where $M$ samples of the random process $U_{k,i}$, for $i\in\left\{ 1,\ldots,M\right\} $,
are measured during the slot $k$ and each measurement $R_{k,i}$
is affected by Additive White Gaussian Noise (AWGN) $Z_{k}$ with
unitary variance. Parameter $Q_{k}$ represents the observation SNR
in slot $k$. From \cite{HE06}, an approximated and analytically
tractable model for $f^{q}\left(D_{k},T_{s,k}\right)$ is\begin{equation}
f^{q}\left(D_{k},T_{s,k}\right)=\frac{N}{b}\times f_{1}^{q}\left(D_{k}\right)\times f_{2}\left(T_{s,k}\right),\label{eq:f_1}\end{equation}
where $b=N/M$ is the bandwidth ratio and $f_{2}\left(T_{s,k}\right)=\zeta\times\max[\left(bT_{s,k}/T_{s}^{max}\right)^{-1/\eta},1]$
models the rate-energy trade-off at the source encoder. The parameter
$\zeta>1$ is related to the efficiency of the encoder, the coefficient
$1\leq\eta\leq3$ is specified by the the given processor \cite{BUR96}
and parameter $T_{s}^{max}$ upper bounds the energy $T_{s,k}$ that
can be used by the source encoder. Function $f_{1}^{q}\left(D_{k}\right)$
is a classical rate-distortion function \cite{COV}. For the model
described in this example, assuming that the source is independent
identically distributed (i.i.d) in time with $U_{k,i}\sim\mathcal{N}\left(0,d_{max}\right)$
, the rate-distortion trade-off is given by\begin{equation}
f_{1}^{q}\left(D_{k}\right)=\left(\log\frac{d_{max}-d_{mmse}}{D_{k}-d_{mmse}}\right)^{+},\mbox{ where }d_{mmse}=\left(\frac{1}{d_{max}}+q\right)^{-1},\label{eq:noisy_est}\end{equation}
 where $d_{mmse}$ is the estimation minimum MSE (MMSE) for the estimate
of $U_{k,i}$ given $R_{k,i}$ \cite{BER71}. Notice that the distortion
$D_{k}$ is upper bounded by $d_{max}$ and lower bounded by $d_{mmse}$.
\begin{example}
The sensor observes \emph{$M$} samples of a first-order Gaussian
Markov source\linebreak$[U_{k,1},U_{k,2},...,U_{k,M}]\in\mathbb{R}^{M}$
with correlation function given by $\mathbb{E}[U_{k}U_{k+j}]=d_{max}Q_{k}^{\left|j\right|}$,
where parameter $0\leq Q_{k}\leq1$ is the correlation coefficient
between the samples measured in slot $k$. Notice that the larger
$Q_{k}$ is, the easier it is for the compressor to reduce the bit
rate for a given distortion due to the increased correlation. Adapting
results from \cite{LU03}%
{}, if the source encoder uses a transform encoder \cite{GER92}, the
optimal compressor produces a number of bits equal to\begin{equation}
f^{q}\left(D_{k},T_{s,k}\right)=\frac{N}{b}\times\left[f_{1}\left(D_{k}\right)+f_{2}^{q}\left(T_{s,k}\right)\right]^{+},\label{eq:f_1-2}\end{equation}
 with $f_{1}\left(D_{k}\right)=\log\frac{\zeta d_{max}}{D_{k}}$,
where parameter $\zeta\geq1$ depends on the type of quantizer, and
\begin{equation}
f_{2}^{q}\left(T_{s,k}\right)=\log\left(1-q^{2}\right)\times\frac{T_{s,k}-\nu/b}{T_{s,k}},\label{eq:no_noise}\end{equation}
 with given parameter $\nu$, which sets a lower bound on the energy
$T_{s,k}$ as $T_{s,k}\geq\nu/b$. Equations (\ref{eq:f_1-2})-(\ref{eq:no_noise})
are obtained by assuming, similar to \cite{LU03}, that the energy
required for source compression is proportional to the size of the
 transform encoder. Finally, notice that, since the compression rate
must be positive, $D_{k}$ is upper bounded by $\zeta d_{max}\left(1-q^{2}\right)^{\frac{T_{s,k}-\nu/b}{T_{s,k}}}$
.
\end{example}
\end{example}

\subsection{Problem Definition\label{sub:Policy Definition}}

At each time slot $k$, a \textit{resource manager} must determine
the distortion $D_{k}$ and the energies $T_{s,k}$ and $T_{t,k}$
to be allocated to the source and channel encoder, respectively. The
decision is taken according to a policy $\pi:=\left\{ \pi_{k}\right\} _{k\geq1}$,
where $\pi_{k}:=\left\{ D_{k}\left(S^{k}\right),T_{s,k}\left(S^{k}\right),T_{t,k}\left(S^{k}\right)\right\} $
determines parameters $(D_{k},T_{s,k},T_{t,k})$ as a function of
the present and past states $S^{k}=\left\{ S_{1},\ldots,S_{k}\right\} $
of the system, where the $S_{i}=\{\tilde{E}_{i},\tilde{X_{i}},Q_{i},H_{i}\}$
accounts for the state of the available energy $\tilde{E}_{i}$ ,
for the data buffer $\tilde{X}_{i}$ , for the the source observation
state $Q_{i}$ and the channel state $H_{i}$. We define the set of
all policies as $\Pi$. Policies can be optimized according to different
criteria. In Sec. \ref{sec:Stability} we adopt stability under an
average distortion criterion as criterion of interest, while Sec.
\ref{sec:Delay-Distortion-Optimization} addresses the optimization
of the trade-off between distortion and delay.

\section{Stability Under a Distortion Constraint\label{sec:Stability}}

In this section, we adopt as performance criterion the stability of
the data queue connecting source and channel encoders. We also impose
the constraint that the policy guarantees the following condition
on the long-term distortion:\begin{equation}
\lim_{n\rightarrow\infty}\frac{1}{n}\sum_{k=1}^{n}D_{k}\leq\bar{D}\label{eq:av_dist}\end{equation}
for a fixed maximum average distortion level $\bar{D}$ tolerated
by the system. We define a policy as $\bar{D}$\textit{-feasible}
if it guarantees the stability of the data queue connecting source
and channel encoders under the average distortion constraint (\ref{eq:av_dist}).
Recall that stability of the data queue holds if the distribution
of $\tilde{X}_{k}$ is asymptotically \textit{stationary} and \textit{proper},
i.e., $\Pr(\tilde{X}_{k}=\infty)\rightarrow0$ \cite{BOR76}.

\subsection{Distortion-Optimal Energy-Neutral Class of Policies\label{sub:Throughput-distortion-optimality}}

For a given distortion $\bar{D}$, our objective in this section is
to identify a class of policies that is able to stabilize the data
queue and satisfy the distortion constraint (\ref{eq:av_dist}) as
long as this is possible. We refer to this class of policies as \textit{distortion-optimal
energy-neutral}. Notice that this definition generalizes that of {}``throughput
optimal'' policies \cite{TAS92} considered in related works such
as \cite{Sha10}, where only the stability constraint is imposed.
By definition, a \textit{distortion-optimal energy-neutral class }\textit{\emph{of
policies}}\textit{ }$\Pi^{do}\subseteq\Pi$ contains at least one
$\bar{D}$-feasible policy. For instance, the set $\Pi$ of all policies
is clearly distortion-optimal energy-neutral. However, this is a rather
unsatisfying solution to the problem. In fact, it does not help in
any way to identify a $\bar{D}$-feasible policy for a given system
setup. Instead, we want to identify a smaller class $\Pi^{do}$, which
is parametrized in a way that makes it easy to evaluate a $\bar{D}$-feasible
policy. The propositions below identify a distortion-optimal energy-neutral
class of policies for the separate source and channel encoders model
depicted in Fig. \ref{Flo:model} and described in Sec. \ref{sec:System-Model}.
\begin{prop}
\label{pro:necessary-conditionTDO}For a given distortion $\bar{D}$,
a necessary condition for the existence of a $\bar{D}$-feasible policy
is the existence of a set of parameters $D^{q}\geq0$, $T_{s}^{q}\geq0$
for $q\in\mathcal{Q}$, $T_{t}^{h}\geq0$ for $h\in\mathcal{H}$,
and $0<\alpha<1$ such that\begin{equation}
\sum_{q}\Pr(q)f^{q}\left(D^{q},T_{s}^{q}\right)<\sum_{h}\Pr(h)g^{h}\left(T_{t}^{h}\right),\;\sum_{q}\Pr(q)D^{q}\leq\bar{D},\label{eq:nec_cond}\end{equation}
\begin{equation}
\sum_{q}\Pr(q)T_{s}^{q}\leq\left(1-\alpha\right)\mathbb{E}\left[E_{k}\right],\;\mbox{ and }\;\sum_{h}\Pr(h)T_{t}^{h}\leq\alpha\mathbb{E}\left[E_{k}\right].\label{eq:channel_cond}\end{equation}
\end{prop}
\begin{rem}
\label{rem:TDO}Parameters $D^{q}$, $T_{s}^{q}$ , $T_{t}^{h}$ and
$\alpha$, whose existence is necessary for the existence of a $\bar{D}$-feasible
policy according to Proposition \ref{pro:necessary-conditionTDO},
have a simple interpretation. In particular, $T_{s}^{q}$ , $D^{q}$
can be read as the average energy and distortion that the source encoder
selects when the observation state is $Q_{k}=q$, whereas $T_{t}^{h}$
can be seen as the average energy that channel encoder draws from
the available energy for transmission when the channel state is $H_{k}=h$.
Moreover, condition (\ref{eq:nec_cond})-left is necessary for the
stability of the data queue, condition (\ref{eq:nec_cond})-right
is necessary to satisfy the constraint (\ref{eq:av_dist}), and conditions
(\ref{eq:channel_cond}) are necessary for energy neutrality. This
interpretation will be used below to derive a class of distortion-optimal
energy-neutral policies.\end{rem}
\begin{IEEEproof}
The processes $f\left(D_{k},T_{s,k},Q_{k}\right)$ and $g\left(H_{k},T_{t,k}\right)$
must be asymptotically stationary ergodic for queue (\ref{eq:coda2})
to be asymptotically stationary. Hence, the policy $\pi$ must be
asymptotically stationary. Under this assumption, the necessary condition
for the distribution of $\tilde{X}_{k}$ to be asymptotically proper
is $\mathbb{E}_{\pi}\left[f\left(D_{k},T_{s,k},Q_{k}\right)\right]<\mathbb{E}_{\pi}\left[g\left(H_{k},T_{t,k}\right)\right]$
from standard results on G/G/1 queues (see any reference on queuing
theory, e.g.,\cite[Ch.3]{BOR76}). Notice that the average $\mathbb{E}_{\pi}$,
with respect to the joint distribution of the state variables $E_{k},Q_{k},H_{k}$,
is explicitly dependent on the policy $\pi_{k}$. From this condition,
since $f$ is separately convex in $D_{k},T_{s,k}$ and $g$ is concave
in $T_{t,k}$, we have the following necessary condition $\sum_{q}\Pr(q)f^{q}\left(\mathbb{E}_{\pi}\left[D_{k}\mid Q_{k}=q\right],\mathbb{E}_{\pi}\left[T_{s,k}\mid Q_{k}=q\right]\right)<$\linebreak$<\sum_{h}\Pr(h)g^{h}\left(\mathbb{E}_{\pi}\left[T_{t,k}\mid H_{k}=h\right]\right)$,
where we have used Jensen inequality on both sides. Defining $D^{q}=\mathbb{E}_{\pi}\left[D_{k}\mid Q_{k}=q\right]$,
$T_{s}^{q}=\mathbb{E}_{\pi}\left[T_{s,k}\mid Q_{k}=q\right]$, and
$T_{t}^{h}=\mathbb{E}_{\pi}\left[T_{t,k}\mid H_{k}=h\right]$, the
condition (\ref{eq:nec_cond})-left is then proved. As for (\ref{eq:channel_cond}),
we consider that, from (\ref{eq:coda1}), we must have $\frac{1}{K}\sum_{k=1}^{K}\left(T_{s,k}+T_{t,k}\right)\leq\frac{1}{K}\sum_{k=1}^{K}E_{k}+\frac{\tilde{E}_{0}}{K},\;\mbox{for }K\geq1$,
and the initial state of the energy buffer $\tilde{E}_{0}$. Then,
for a stationary ergodic policy $\pi$, we get $\mathbb{E}_{\pi}\left[T_{s,k}\right]+\mathbb{E}_{\pi}\left[T_{t,k}\right]\leq\mathbb{E}\left[E_{k}\right]$,
where $\frac{1}{K}\sum_{k=1}^{K}T_{s,k}\rightarrow\mathbb{E}_{\pi}\left[T_{s,k}\right]$,
$\frac{1}{K}\sum_{k=1}^{K}T_{t,k}\rightarrow\mathbb{E}_{\pi}\left[T_{t,k}\right]$,
and $\frac{1}{K}\sum_{k=1}^{K}E_{k}+\frac{\tilde{E}_{0}}{K}\rightarrow\mathbb{E}\left[E_{k}\right]$.
Given the definitions and the inequality above, (\ref{eq:channel_cond})
are proved, having set $\alpha=\mathbb{E}_{\pi}\left[T_{t,k}\right]/\mathbb{E}\left[E_{k}\right]$.
To conclude, for (\ref{eq:nec_cond})-right, we observe that the distortion
constraint (\ref{eq:av_dist}) is satisfied.
\end{IEEEproof}
We now look for a distortion-optimal energy-neutral class of policies.
To this end, based on Proposition \ref{pro:necessary-conditionTDO},
it is enough to exhibit a class of policies such that it contains
a $\bar{D}$-feasible policy as long as the necessary conditions (\ref{eq:nec_cond})-(\ref{eq:channel_cond})
are satisfied for some set of parameters $D^{q}$, $T_{s}^{q}$ ,
$T_{t}^{h}$ and $\alpha$. Proposition \ref{pro:necessary-conditionTDO}
suggests that it is possible to find $\bar{D}$-feasible policies
that select $D_{k}$ and $T_{s,k}$ based on the observation state
$Q_{k}$ only, whereas the selection of $T_{t,k}$ depends on the
channel state $H_{k}$ only. Based on this consideration, let us define
the class of policies $\Pi^{do}$\begin{equation}
\Pi^{do}\;\begin{cases}
D_{k}=D^{q},\; T_{s,k}=\min\left[\left(1-\alpha\right)\tilde{E}_{k}-\epsilon,T_{s}^{q}\right] & \mbox{for \ensuremath{Q_{k}=q}}\\
T_{t,k}=\min\left[\alpha\tilde{E}_{k}-\epsilon,T_{t}^{h}\right] & \mbox{for }H_{k}=h\end{cases}\label{eq:policy_TO}\end{equation}
where $D^{q}\geq0$, $T_{s}^{q}\geq0$ for $q\in\mathcal{Q}$, $T_{t}^{h}\geq0$
for $h\in\mathcal{H}$, and $0<\alpha<1$ are fixed design parameters.
\begin{prop}
\label{pro:Sufficient-conditions-TDO}A policy in\textup{ $\Pi^{do}$}
is $\bar{D}$-feasible if conditions (\ref{eq:nec_cond}) hold, along
with\begin{equation}
\sum_{q}\Pr(q)T_{s}^{q}\leq\left(1-\alpha\right)\mathbb{E}\left[E_{k}\right]-\epsilon,\;\mbox{ and }\;\sum_{q}\Pr(q)T_{t}^{h}\leq\alpha\mathbb{E}\left[E_{k}\right]-\epsilon.\label{eq:suff_cond_channel}\end{equation}
\end{prop}
\begin{rem}
\label{rem:The-sufficient-conditionsTDO}The sufficient conditions
in Proposition \ref{pro:Sufficient-conditions-TDO} for the policies
in $\Pi^{do}$ to be $\bar{D}$-feasible coincide, for $\epsilon\rightarrow0$,
with the necessary conditions derived in Proposition \ref{pro:necessary-conditionTDO}.
Therefore $\Pi^{do}$ contains a $\bar{D}$-feasible policy any time
the necessary conditions of Proposition \ref{pro:necessary-conditionTDO}
hold. As discussed above, this implies that the \emph{set $\Pi^{do}$
is a distortion-optimal energy-neutral class}. Moreover, it should
be noted that the class $\Pi^{do}$, given (\ref{eq:policy_TO}),
is parametrized by a small number of parameters and the policies in
$\Pi^{do}$ perform separate resource allocation optimizations for
the source and channel encoders. In particular, the energy allocated
to the source encoder $T_{s,k}$ only depends on the observation state
$Q_{k}$, and not on the channel quality $H_{k}$, whereas the energy
$T_{t,k}$ for the channel encoder only depends on $H_{k}$, and not
on $Q_{k}$. The energy allocation between the two encoders is governed
by a single parameter $0<\alpha<1$. This entails that, once this
parameter is fixed, and thus the energy budget available at the two
encoders is fixed, resource allocation at the two encoders can be
done separately without loss of optimality.\end{rem}
\begin{IEEEproof}
For $0<\alpha<1$ such that $\sum_{q}\Pr(q)T_{s}^{q}\leq\left(1-\alpha\right)\mathbb{E}\left[E_{k}\right]-\epsilon$
and $\sum_{q}\Pr(q)T_{t}^{h}\leq\alpha\mathbb{E}\left[E_{k}\right]-\epsilon$,
with $\epsilon$ small, we obtain that $\Pr(\tilde{E}_{k}=\infty)=1$
asymptotically. This is true since $\mathbb{E}\left[T_{s,k}+T_{t,k}\right]<\mathbb{E}\left[E_{k}\right]$
in the system (\ref{eq:coda1}), so that the energy harvested is larger
than the energy consumed on average and the energy queue is not stable
\cite[Ch.3]{BOR76} (see also \cite{Sha10} for the same argument).
This leads to the asymptotically infinite size of the stored energy,
as the buffer capacity is assumed infinite. Therefore we have, $T_{s,k}\left(Q_{k}=q\right)\rightarrow T_{s}^{q}$
and $T_{t,k}\left(H_{k}=h\right)\rightarrow T_{t}^{h}$ from (\ref{eq:policy_TO}).
Notice that this argument shows that in (\ref{eq:policy_TO}) one
can substitute $\alpha$ for any number between 0 and 1 leading to
the same sufficient conditions (\ref{eq:nec_cond}) and (\ref{eq:suff_cond_channel}).
Due to the stationarity and ergodicity of the processes $Q_{k}$ and
$H_{k}$, $f\left(D_{k},T_{s,k},Q_{k}\right)$ and $g\left(H_{k},T_{t,k}\right)$
are stationary ergodic, and $\sum_{q}\Pr(q)f^{q}\left(D^{q},T_{s}^{q}\right)<\sum_{h}\Pr(h)g^{h}\left(T_{t}^{h}\right)$
is the sufficient condition for the stability of the queue $\tilde{X}_{k}$
\cite[Ch.3]{BOR76}. Being $\lim_{n\rightarrow\infty}\frac{1}{n}\sum_{k=1}^{n}D_{k}=\sum_{q}\Pr(q)D^{q}$,
for $D^{q}$ such that $\sum_{q}\Pr(q)D^{q}\leq\bar{D}$ the class
of policies $\Pi^{do}$ satisfies the constraint (\ref{eq:av_dist}). \end{IEEEproof}
\begin{rem}
A problem of interest is to find the minimal distortion $\bar{D}$
for which the set of distortion-optimal energy-neutral policies $\Pi^{do}$
is not empty. In other words, assessing the minimal distortion that
can be supported without causing the data queue to be unstable. Given
the separate nature of the source and channel energy allocations,
it can be seen that one should optimize both terms in (\ref{eq:nec_cond})-left
separately, once the optimal value for $\alpha$ has been found. In
particular, when $g\left(H_{k},T_{t,k}\right)$ is the Shannon capacity,
the policy $T_{t,k}$ that minimizes $\bar{D}$ is the \textit{water-filling}
\cite{COV}.
\end{rem}

\subsection{Suboptimal Classes of Policies\label{sub:Suboptimal-policies}}

In Sec. \ref{sub:Throughput-distortion-optimality}, a distortion-optimal
energy-neutral class $\Pi^{do}$ has been identified. This class of
policies, as made clear by the proof of Proposition \ref{pro:Sufficient-conditions-TDO}
requires infinite energy storage capabilities at the sensor node.
Let us instead consider the class of greedy policies $\Pi^{sub1}$
that do not use the energy buffer but allocates all the energy arrival
$E_{k}$ to source and channel coding according to a fraction $0\leq\alpha^{q,h}\leq1$
that depends on both source $Q_{k}=q$ and channel $H_{k}=h$ states:\begin{equation}
\Pi^{sub1}\;\begin{cases}
D_{k}=D^{q,h},\; T_{s,k}=\alpha^{q,h}E_{k},\; T_{t,k}=(1-\alpha^{q,h})E_{k} & \mbox{for \ensuremath{Q_{k}=q\mbox{ and }H_{k}=h}}\end{cases}\label{eq:policy_sub}\end{equation}
where the distortion $D^{q,h}\geq0$ also depends on both source and
channel states. Notice that this is unlike the class of distortion-optimal
energy-neutral policies (\ref{eq:policy_TO}) in which, as explained
in Remark \ref{rem:The-sufficient-conditionsTDO}, energy allocation
is done independently for source (only based on $Q_{k}$) and channel
decoder (only based on $H_{k}$). Here, parameters $\alpha^{q,h},D^{q,h}$
are selected on the basis of both channel and source states $Q_{k}$
and $H_{k}$ to partially compensate for the loss due to the greedy
approach. For further reference, we also consider the subclass of
policies $\Pi^{sub2}:=\Pi^{sub1}|_{\alpha^{q,h}=\alpha\mbox{ for all }\left(q,h\right)}$,
for which the power allocation is not adapted to the channel and observation
states.
\begin{prop}
\label{pro:Sufficient-conditions-greedy}Policies in $\Pi^{sub1}$
are $\bar{D}$-feasible if the following conditions hold: \begin{equation}
\sum_{q}\sum_{h}\Pr(q)\Pr(h)\mathbb{E}\left[f^{q}\left(D^{q,h},\alpha^{q,h}E_{k}\right)\right]<\sum_{q}\sum_{h}\Pr(q)\Pr(h)\mathbb{E}\left[g^{h}\left((1-\alpha^{q,h})E_{k}\right)\right],\label{eq:cond_suff_sub}\end{equation}
\textup{\begin{equation}
\mbox{ and }\sum_{q}\sum_{h}\Pr(q)\Pr(h)D^{q,h}\leq\bar{D},\label{eq:cond_suff_sub_dist}\end{equation}
where the expectation $\mathbb{E}$ in (\ref{eq:cond_suff_sub}) is
over the energy harvesting process $E_{k}$.}\end{prop}
\begin{rem}
\label{rem:greedy}In general, the set of policies $\Pi^{sub1}$ is
not guaranteed to be a distortion-optimal energy-neutral class, since
the necessary conditions of Proposition \ref{pro:necessary-conditionTDO}
could hold where the sufficient conditions of Proposition \ref{pro:Sufficient-conditions-greedy}
do not. This is also confirmed via numerical simulations in Sec. \ref{sub:Stability-regions:-numerical}.
However, for constant observation and channel states, i.e., $H_{k}=h_{0}$
and $Q_{k}=q_{0}$ for all $k$, and for $f$ and $g$ linear in $T_{s,k}$
and $T_{t,k}$, respectively, the class $\Pi^{sub1}$ is distortion-optimal
energy-neutral. In fact, under these assumptions, the sufficient condition
(\ref{eq:cond_suff_sub}) becomes $f^{q_{0}}\left(D^{q_{0},h_{0}},\alpha^{q_{0},h_{0}}\mathbb{E}\left[E_{k}\right]\right)<g^{h_{0}}\left(\left(1-\alpha^{q_{0},h_{0}}\right)\mathbb{E}\left[E_{k}\right]\right)$.
Defining $T_{s}^{q}=\alpha^{q_{0},h_{0}}\mathbb{E}\left[E_{k}\right]$,
$T_{t}^{h}=\left(1-\alpha^{q_{0},h_{0}}\right)\mathbb{E}\left[E_{k}\right]$
and $D^{q}=D^{q_{0},h_{0}}$, conditions (\ref{eq:cond_suff_sub})-(\ref{eq:cond_suff_sub_dist})
correspond to (\ref{eq:nec_cond}). Thus, the class of policies $\Pi^{sub1}$
is distortion-optimal energy-neutral.\end{rem}
\begin{IEEEproof}
Due to the stationarity and ergodicity of the processes $Q_{k}$ and
$H_{k}$, the parameters $D_{k}$, $T_{s,k}$ and $T_{t,k}$, are
also stationary ergodic and (\ref{eq:cond_suff_sub}) is a sufficient
condition for the stability of the queue $\tilde{X}_{k}$ \cite[Ch.3]{BOR76}.
Since $\lim_{n\rightarrow\infty}\frac{1}{n}\sum_{k=1}^{n}D_{k}=\sum_{q}\sum_{h}\Pr(q)\Pr(h)D^{q,h}$,
for $D^{q,h}$ such that $\sum_{q}\sum_{h}\Pr(q)\Pr(h)D^{q,h}\leq\bar{D}$
the class of policies $\Pi^{sub1}$ satisfies the constraint (\ref{eq:av_dist}).
\end{IEEEproof}
The greedy policies introduced above do not make use of the energy
buffer at all, whereas the distortion-optimal energy-neutral class
of policies $\Pi^{do}$ does. For comparison purposes, it is interesting
to consider hybrid policies that differ from those in $\Pi^{do}$
as the energy buffer is used only either for compression or for transmission.
The first policies $\Pi^{hyb1}$ require an energy buffer for the
channel encoder only in order to adapt the transmission power to the
channel state, i.e., $T_{t,k}=\min[\alpha\tilde{E}_{k},T_{t}^{h}]\mbox{ for }H_{k}=h$.
The energy allocated to the source encoder is instead independent
of the observation state, i.e., $T_{s,k}=\left(1-\alpha\right)E_{k}$.
Viceversa, the second policies $\Pi^{hyb2}$ are adapted to the observation
state instead of the channel state, and require an energy buffer only
for the source encoder, i.e., $T_{s,k}=\min[\left(1-\alpha\right)\tilde{E}_{k},T_{s}^{q}]\mbox{ for \ensuremath{Q_{k}=q}}$
and $T_{t,k}=\alpha E_{k}$.
\begin{prop}
\label{pro:Sufficient-conditions-hybrids}Policies in $\Pi^{hyb1}$
are $\bar{D}$-feasible if conditions (\ref{eq:nec_cond})-right and
(\ref{eq:suff_cond_channel})-right hold, along with\textup{ $\sum_{q}\Pr(q)\mathbb{E}\left[f^{q}\left(D^{q},\left(1-\alpha\right)E_{k}\right)\right]<\sum_{h}\Pr(h)g^{h}\left(T_{t}^{h}\right)$}.
Similarly, policies $\Pi^{hyb2}$ are $\bar{D}$-feasible if conditions
(\ref{eq:nec_cond})-right and (\ref{eq:suff_cond_channel})-left
hold, along with $\sum_{q}\Pr(q)f^{q}\left(D^{q},T_{s}^{q}\right)<$\linebreak$<\sum_{h}\Pr(h)\mathbb{E}\left[g^{h}\left(\alpha E_{k}\right)\right]$\textup{.} \end{prop}
\begin{IEEEproof}
Proof follows from the proofs of Propositions \ref{pro:Sufficient-conditions-TDO}
and \ref{pro:Sufficient-conditions-greedy}.
\end{IEEEproof}

\subsection{Analog Transmission\label{sub:Analog-transmission}}

In this section, we consider for performance comparison an alternative
class of strategies in which the sampled source is transmitted directly
via analog modulation (see, e.g., \cite{GAS03}). In other words,
a block of source samples is scaled and transmitted in one slot, so
as to consume $T_{t,k}$ transmission energy per channel use during
slot $k$. Energy $T_{t,k}$ is selected as $T_{t,k}=\min[\tilde{E}_{k}-\epsilon,T_{t}^{q,h}]\mbox{ for }Q_{k}=q\mbox{ and }H_{k}=h$
for given parameters $T_{t}^{q,h}\geq0$, so that it depends on the
current source and channel states. If the bandwidth ratio $b=N/M$
is larger than one, i.e., there are more channel uses than source
samples, the extra $N-M$ source samples are unused. Instead, if $b<1$,
then a fraction $1-b$ of source samples is not transmitted. Notice
that this class of strategies does not fall in the category depicted
in Fig. \ref{Flo:model} and discussed above, since it does not have
separate encoders.

We assume the observation model of Example 1 with an i.i.d. source
$U_{k,i}\sim\mathcal{N}\left(0,d_{max}\right)$ and an AWGN channel
with SNR $H_{k}$. For a bandwidth ratio of $b$, it is not difficult
to obtain that the MMSE at the receiver is given by \begin{equation}
d_{mmse}\left(T_{t,k},Q_{k},H_{k}\right)=\;\begin{cases}
\left(\frac{bT_{t,k}Q_{k}H_{k}}{bT_{t,k}Q_{k}+Q_{k}+1}+\frac{1}{d_{max}}\right)^{-1} & \mbox{for }b\geq1\\
b\times\left(\frac{T_{t,k}Q_{k}H_{k}}{T_{t,k}Q_{k}+Q_{k}+1}+\frac{1}{d_{max}}\right)^{-1}+\left(1-b\right)d_{max} & \mbox{for }b<1\end{cases}.\label{eq:mmse_analog}\end{equation}
Notice that, for fairness, the average energy used for the transmission
of one sample is $bT_{t,k}$ if $b\geq1$. Also, notice that if $b<1$,
the maximum distortion $d_{max}$ is accrued on the fraction $(1-b)$
of samples that are not transmitted. For simplicity, we assume that
analog transmission has negligible power spent for source acquisition,
i.e., $T_{s,k}=0$, though this is not entirely correct given that
even in this case there is a need for sensing, sampling and analog-to-digital
conversions. Nonetheless, these power consumption terms are also neglected
in Examples 1 and 2. Under this assumption, we have the following.
\begin{prop}
\label{pro:nec_suff_anlog}Analog transmission satisfies the distortion
constraint (\ref{eq:av_dist}) if the following conditions are satisfied:\begin{equation}
\sum_{q}\sum_{h}\Pr(q)\Pr(h)d_{mmse}\left(T_{t}^{q,h},Q=q,H=h\right)\leq\bar{D},\;\mbox{ and }\;\sum_{q}\sum_{h}\Pr(q)\Pr(h)T_{t}^{q,h}\leq\mathbb{E}\left[E_{k}\right].\label{eq:analog_nec_cond}\end{equation}
\end{prop}
\begin{IEEEproof}
Follows similar to Proposition \ref{rem:The-sufficient-conditionsTDO}.
\end{IEEEproof}
We can also consider a greedy policy $T_{t,k}=E_{k}$, for which the
power allocation is not adapted to the channel and observation states
and energy storage is not required.

\subsection{Numerical Results\label{sub:Stability-regions:-numerical}}

In this section we compare numerically the performance of the optimal
and suboptimal source-channel coding policies presented so far, along
with analog transmission strategies.

\begin{figure}[t]
\centering{\includegraphics[width=15cm]{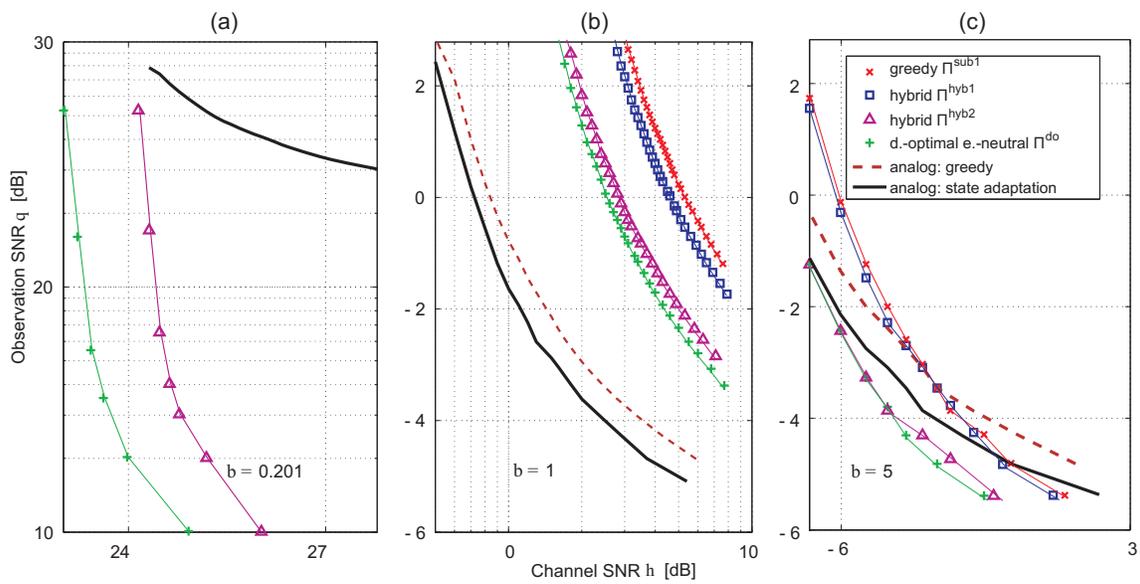}}\caption{Achievable regions (regions above the curves) for the digital policies
(lines with markers) and for the analog policies (only lines): (a)
bandwidth ratio $b=0.201$; (b) bandwidth ratio $b=1$; (c) bandwidth
ratio $b=5$ ($E_{k}\sim\mathcal{U}\left(0,2\right)$, with mean 1
Joule/channel use; $\bar{D}=0.8$; compression model (\ref{eq:f_1}),
with $T_{s}^{max}=1$ Joule/source sample, $\zeta=1$, $\eta=1.5$,
maximum distortion $d_{max}\mbox{=}1$).}

\label{Flo:stabilities_one_state}%
\end{figure}

Consider first a scenario where the observation and channel states
are constant, i.e., $Q_{k}=q$ and $H_{k}=h$ for all $k$. The energy
arrival $E_{k}$ has mean 1 Joule/channel use and uniform pdf between
0 and 2 Joule/channel use. We consider model (\ref{eq:f_1}) with
$T_{s}^{max}=1$ Joule/source sample, efficiency parameters $\zeta=1$
and $\eta=1.5$ for the source encoder, and the complex AWGN channel
Shannon capacity $g^{h}\left(T_{t,k}\right)=N\times\log(1+hT_{t,k})$.
In Fig. \ref{Flo:stabilities_one_state}, we identify the values of
source and channel SNRs ($q,h$) for which different policies are
able to stabilize the data queue and guarantee average distortion
$\bar{D}=0.8$. We refer to these regions as {}``achievable regions''.
Achievable regions are given in Fig. \ref{Flo:stabilities_one_state}
by the area above the corresponding lines. We use standard tools of
convex optimization for their numerical evaluation.

In Fig. \ref{Flo:stabilities_one_state}-b, we can further observe
that the achievable regions of the distortion-optimal energy-neutral
class (\ref{eq:policy_TO}) are significantly larger than those of
the greedy policies (\ref{eq:policy_sub}) due to possibility to store
energy and thus allocate resources more effectively. Moreover, by
considering also the hybrid policies $\Pi^{hyb1}$ and $\Pi^{hyb2}$
, we can see that most of the gains are obtained, in this example,
by exploiting the energy buffer in order to allocate energy over time
to the source encoder, whereas the gains accrued by using the battery
for data transmission are less significant. This is observed by noticing
that the achievable region of the class $\Pi^{do}$ is close to that
obtained by hybrid policies $\Pi^{hyb2}$, but much larger than that
obtained by hybrid class of policies $\Pi^{hyb1}$. The relative comparison
between the two hybrid policies, and thus between the use of the battery
for source or channel encoding, depends on the functions $f\left(D_{k},T_{s,k},Q_{k}\right)$
and $g\left(H_{k},T_{t,k}\right)$. For instance, setting a lower
$T_{s}^{max}$ would change the presented results by penalizing more
the strategies that are not using the energy buffer for channel transmission.

We now consider the performance of analog transmission. As it is well
known from rate-distortion theory \cite{BER71}, for bandwidth ratio
$b=1$ and the given (Gaussian) source and channel models, analog
transmission is rate-distortion optimal. Separate source-channel coding
is also optimal (for any $b>0$) if compression is assumed not to
consume any energy. Here, instead, the achievable region of analog
transmission is expected to be larger than that of strategies that
employ separate source-channel coding, as source encoding energy costs
are taken into account in the given model (\ref{eq:f_1}). This is
confirmed by Fig. \ref{Flo:stabilities_one_state}-b. However, for
sufficiently larger or smaller bandwidth ratios, the extra energy
spent for compression is not enough to overcome the rate-distortion
gains attained by separate source-channel coding versus analog transmission.
This is apparent from Fig. \ref{Flo:stabilities_one_state}-a and
Fig. \ref{Flo:stabilities_one_state}-c where the analog transmission
is outperformed.

\begin{figure}[t]
\centering{\includegraphics[width=16cm]{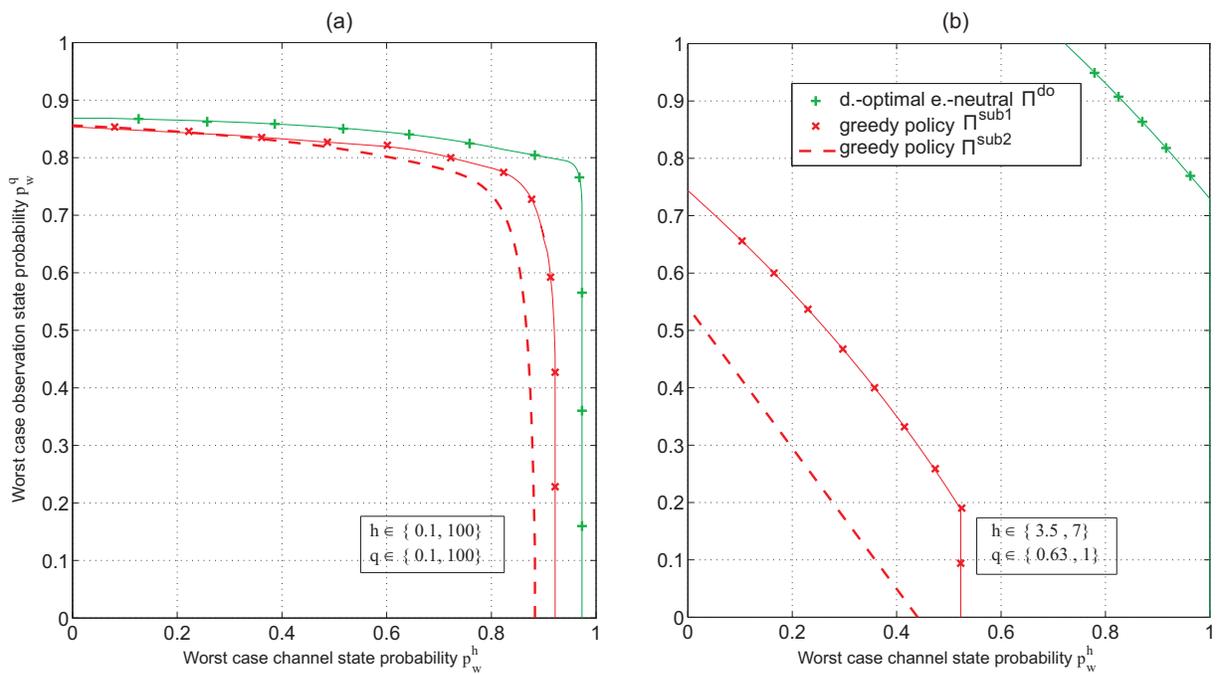}}\caption{Achievable regions (regions below the curves) of the digital policies
$\Pi^{do}$ (\ref{eq:policy_TO}), $\Pi^{sub1}$ (\ref{eq:policy_sub}),
and $\Pi^{sub2}$, with two channel and observation SNR states, respectively.
($E_{k}\sim\mathcal{U}\left(0,2\right)$, with mean 1 Joule/channel
use; $\bar{D}=0.8$; compression model (\ref{eq:f_1}), with $T_{s}^{max}=1$
Joule/source sample, $\zeta=1$, $\eta=1.5$, bandwidth ratio $b=1$,
maximum distortion $d_{max}\mbox{=}1$).}

\label{Flo:stab_two_states}%
\end{figure}

We now consider a scenario where source and channel states are not
constant but vary with two possible states, namely $\mathcal{Q}=\left\{ 10^{-1},10^{2}\right\} $
(Fig. \ref{Flo:stab_two_states}-a) or $\mathcal{Q}=\left\{ 10^{-0.2},1\right\} $
(Fig. \ref{Flo:stab_two_states}-b) for source SNR, and $\mathcal{H}=\left(10^{-1},10^{2}\right)$
(Fig. \ref{Flo:stab_two_states}-a) or $\mathcal{H}=\left(3.5,7\right)$
(Fig. \ref{Flo:stab_two_states}-b) for channel SNR. The worst-case
observation SNR (e.g., $Q_{k}=10^{-1}$ for Fig. \ref{Flo:stab_two_states}-a)
and the worst-case channel SNR (e.g., $H_{k}=10^{-1}$ for Fig. \ref{Flo:stab_two_states}-a)
have probabilities $p_{w}^{q}$ and $p_{w}^{h}$, respectively. In
Fig. \ref{Flo:stab_two_states}, achievability regions are the sets
of probability values $(p_{w}^{q},p_{w}^{h})$ for which different
policies guarantee queue stability and average distortion $\bar{D}=0.8$.
In particular, the regions are identified as the area below the corresponding
curves. The results emphasize the importance of jointly adapting the
resource allocation to both source and channel states in case of a
greedy policy that does not employ the battery. This is seen by comparing
the performance of the greedy schemes $\Pi^{sub1}$ (\ref{eq:policy_sub}),
which adapts the policy to the current states, and $\Pi^{sub2}$,
which does not. Moreover, comparing Fig. \ref{Flo:stab_two_states}-b
with Fig. \ref{Flo:stab_two_states}-a, it is seen that the better
{}``worst-case'' state allows the distortion-optimal energy-neutral
policy to satisfy stability and average distortion constraints for
larger values of the probabilities $(p_{w}^{q},p_{w}^{h})$. On the
contrary, the greedy policies suffer from the worse {}``best-case''
state $(q,h)=(1,7)$ of Fig. \ref{Flo:stab_two_states}-b, as this
corresponds to operating critically close to the border of their achievable
regions (see the achievable region of $\Pi^{sub1}$ in Fig. \ref{Flo:stabilities_one_state}-b).

\section{Delay-Distortion Optimization\label{sec:Delay-Distortion-Optimization}}

The stability criterion considered in Sec. \ref{sec:Stability} does
not provide any guarantee on the delay experienced by the reconstruction
of the source in a certain time-slot. In some applications, instead,
one may be willing to trade distortion for a shorter delay. In this
section, we address such requirement by looking for policies that
minimize a weighted sum of distortion and delay. In particular, we
propose to minimize the \textit{expected total discounted cost }\cite{Put}\begin{equation}
\lim_{n\rightarrow\infty}\frac{1}{n}\sum_{k=0}^{n}\lambda^{k}\left[\gamma D_{k}+\left(1-\gamma\right)\tilde{X}_{k}\right],\label{eq:delay_dist}\end{equation}
where $0\leq\lambda<1$ is the discount factor and $0\leq\gamma\leq1$.
The latter parameter weights the importance of distortion versus delay
in the optimization criterion. Notice that if $\gamma=0$ then one
minimizes the average length of the data queue, which, by Little's
theorem, is the same as minimizing the average delay.

In order to tackle the minimization of (\ref{eq:delay_dist}) over
the policies $\pi$ defined in Sec. \ref{sec:Stability}, we assume
that: (\emph{i}) \textit{\emph{The data and energy buffers are finite;
}}(\emph{ii}) \textit{\emph{The set of possible decisions }}\textit{$\pi_{k}:=\left\{ D_{k},T_{s,k},T_{t,k}\right\} $
}\textit{\emph{is discrete; (}}\textit{iii}\textit{\emph{)}} \textit{\emph{The
energy arrival }}\textit{$E_{k}$ }\textit{\emph{takes values in a
discrete and finite set}};\textit{\emph{ (}}\textit{iv}\textit{\emph{)
The sets of values assumed by rates }}$f\left(D_{k},T_{s,k},Q_{k}\right)$,
$g\left(H_{k},T_{t,k}\right)$, and by\textit{\emph{ the queue length
$\tilde{X_{k}}$ are discrete. Following }}standard theory \cite[Ch. 6]{Put},
these assumptions entail that the optimal policy is deterministic
and stationary (Markovian). In other words, $\left\{ D_{k},T_{s,k},T_{t,k}\right\} $
are function of the present state $S_{k}=\{\tilde{E}_{k},\tilde{X_{k}},Q_{k},H_{k}\}$
only. Therefore, the solution can be found via value iteration \cite{Put}.
Notice that, due to (\emph{i}), data buffer overflow may happen, in
which case the compression bits are lost and a maximum distortion
$d_{max}$ is accrued for the current slot.

While in general the optimal policy allocates resources to source
and channel encoder through parameters $T_{t}^{q,h}$ as a function
of both source $Q_{k}=q$ and channel $H_{k}=h$ states, the class
of policies $\Pi^{do}$ (\ref{eq:policy_TO}) performs such allocation
independently for source and channel encoders. For comparison purposes,
we evaluate also the performance in terms of criterion (\ref{eq:delay_dist})
of a class of policies that optimize separately the source encoder
parameters $\left\{ D_{k},T_{s,k}\right\} $ as a function of $Q_{k}$,
and the channel encoder parameter $\left\{ T_{t,k}\right\} $ as a
function of $H_{k}$. As for the distortion-optimal energy-neutral
class of policies, the energy resources are split between the encoders,
such that the source encoder makes use of a fraction $\alpha$ of
the energy, whereas the rest is utilized by the channel encoder. Specifically,
the energy-buffer is divided into two buffers, that are used independently
by the encoders: the source encoder buffer is charged by $\alpha E_{k}$,
while the channel encoder buffer absorbs the remaining quantity of
energy arrival $\left(1-\alpha\right)E_{k}$. The source encoder policy
$\left\{ D_{k},T_{s,k}\right\} $ is optimized via value iteration
with respect to the criterion (\ref{eq:delay_dist}) by assuming a
constant transmission rate $g\left(H_{k},T_{t,k}\right)=\bar{g}$.
On the other hand, the channel encoder policy $\left\{ T_{t,k}\right\} $
is optimized via value iteration with respect to criterion (\ref{eq:delay_dist})
with weight $\gamma=0$ (since it cannot optimize its policy with
respect to the distortion), by assuming a constant source rate $f\left(D_{k},T_{s,k},Q_{k}\right)=\bar{f}$.
The best {}``separable'' policy is finally obtained by selecting
the values $\left(\alpha,\bar{g},\bar{f}\right)$ that achieve the
best delay-distortion trade-off. %
{}

\subsection{Numerical Results}

In this section we compute numerically the trade-off between delay
and distortion by minimizing (\ref{eq:delay_dist}) for different
values of $\gamma$. Specifically, for each $\gamma$, we evaluate
the average delay, which is measured by the average data queue length
by Little's law, and average distortion. The optimal policies are
computed via value iteration \cite{Put} and so are the suboptimal
policies corresponding to separate optimization of source and channel
encoders.

\begin{figure}[t]
\centering{\includegraphics[width=16cm]{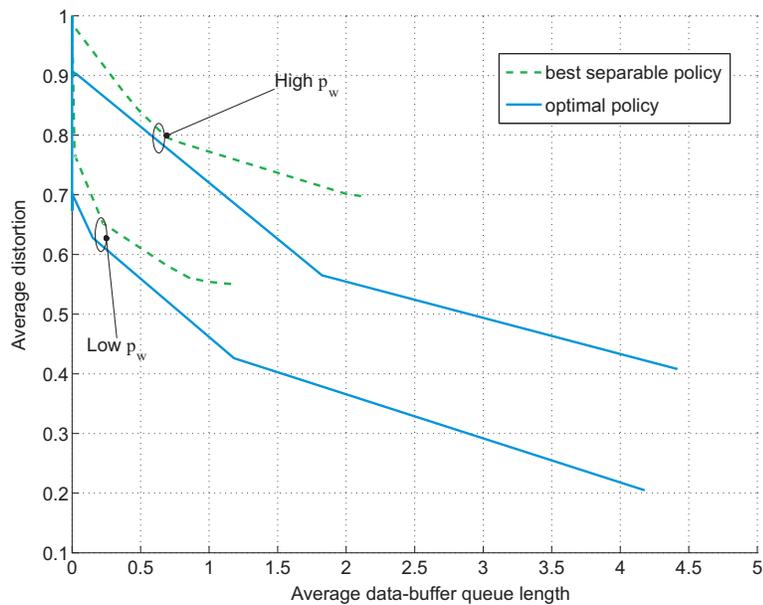}}\caption{Delay-distortion trade-off, where average delay is proportional to
the depicted average data queue length (maximum data-buffer length:
5 codeword lengths; maximum distortion: $d_{max}\mbox{=}1$; discount
factor: $\lambda=0.5$; compression model (\ref{eq:f_1-2}), with
minimum required energy per sample $\nu=0.1$ Joule/sample; bandwidth
ratio: $b=1$; queue length: $\tilde{X}_{k}\in\left\{ 0,\ldots,5\right\} $
expressed in multiples of $M$; energy buffer size: $\tilde{E}_{k}-E_{k}\in\left\{ 0,1,2\right\} $;
energy arrival: $E_{k}\in\left\{ 1,2\right\} $ Joule/sample; source
correlation values: $\mathcal{Q}=\left\{ 0.1,0.5\right\} $; channel
SNR values: $\mathcal{H}=\left(0.5,10\right)$; distortion values:
$D_{k}\in\left\{ 0.1,0.55,1\right\} $). }

\label{Flo:delay_dist}%
\end{figure}

In Fig. \ref{Flo:delay_dist} the delay-distortion trade-off is shown
both for the optimal policies and for the {}``separable'' ones discussed
above. The discount factor is $\lambda=0.5$. The compression model
is (\ref{eq:f_1-2}), with minimum required energy per sample $\nu=0.1$
Joule/sample and bandwidth ratio $b=1$. The quantities of interest
are discretized as follows: $\tilde{X}_{k}\in\left\{ 0,\ldots,5\right\} $
is expressed in multiples of the codeword length $M=N$; The energy
buffer size is $\tilde{E}_{k}-E_{k}\in\left\{ 0,1,2\right\} $ and
$E_{k}\in\left\{ 1,2\right\} $ with $p_{w}^{e}$ being the probability
that $E_{k}=1$ (worst case); The source correlation values are $\mathcal{Q}=\left\{ 0.1,0.5\right\} $
and channel SNR values are $\mathcal{H}=\left(0.5,10\right)$, with
probabilities $p_{w}^{q}$ and $p_{w}^{h}$ for $Q_{k}=0.1$ and $H_{k}=0.5$
(worst cases); The distortion takes values as $D_{k}\in\left\{ 0.1,0.55,1\right\} $
and $d_{max}=1$; The source-encoder rate $f\left(D_{k},T_{s,k},Q_{k}\right)/M$
is rounded to the smallest following integer, while the channel-encoder
rate $g\left(H_{k},T_{t,k}\right)/N$ is rounded to the largest previous
integer.

In Fig. \ref{Flo:delay_dist} we observe that the optimal policies
obtain a remarkably better delay-distortion trade-off compared to
the separable policies, both for low and large worst-case probabilities,
i.e., $p_{w}^{e}=p_{w}^{q}=p_{w}^{h}=p_{w}=0.1\mbox{ and }0.9$. This
demonstrates the importance of a joint resource allocation over the
encoders whenever the delay is also of interest. Note that for increasing
average buffer length (delay), since the buffer size is finite, it
becomes more crucial to adopt a joint resource allocation. This is
because the separate approach is not able to prevent buffer overflow
as the source encoder operates without channel state information.

\section{Multiple Access\label{sec:Scheduling-Protocols}}

In this section, we briefly discuss an extension of the analysis to
a scenario in which $L$ sensors access a single access point employing
TDMA. Random access protocols will be considered in future work (an
analysis with exogenous bit arrivals, and thus no source encoder,
can be found in \cite{Sha09}). Each sensor is modeled as described
in Sec. \ref{sec:System-Model}, and we assume that observation qualities
$\mathbf{Q}_{k}=\left[Q_{1,k},\ldots,Q_{L,k}\right]^{T}\in\mathcal{Q}^{L}=\left[\mathcal{Q}_{1},\ldots,\mathcal{Q}_{L}\right]$,
channel qualities $\mathbf{H}_{k}=\left[H_{1,k},\ldots,H_{L,k}\right]^{T}\in\mathcal{H}^{L}=\left[\mathcal{H}_{1},\ldots,\mathcal{H}_{L}\right]$,
and the energy arrivals $\mathbf{E}_{k}\in\mathbb{R}_{+}^{L}$ are
jointly stationary and ergodic. We tackle the problem of designing
policies defined, extending Sec. \ref{sub:Policy Definition}, as
the tuple $\upsilon:=\left\{ \tau_{k},\boldsymbol{\pi}_{k}\right\} _{k\geq1}$
that consists of the scheduling policy $\tau_{k}\in\left\{ 1,\ldots,L\right\} $,
which reserves the slot $k$ to one sensor $l\in\left\{ 1,\ldots,L\right\} $,
and of the joint resource allocation policy $\mathbf{\boldsymbol{\pi}}_{k}=\left[\pi_{1,k},\ldots,\pi_{L,k}\right]^{T}$,
where each entry $\pi_{l,k}$ is defined as in Sec. \ref{sub:Policy Definition}
and corresponds to the distortion and energy allocation for the $l^{th}$
sensor. Recall that $\left\{ \tau_{k},\boldsymbol{\pi}_{k}\right\} $
generally depends on the whole history of past and current states
(see Sec. \ref{sub:Policy Definition}). Note that time-slot $k$
is exclusively assigned to sensor $l,$ i.e., $T_{t,l,k}=0$ if $\tau_{k}\neq l$.
We define a policy $\upsilon$ as $\mathbf{\bar{D}}$-feasible if
it guarantees the stability of all data queues and average distortion
constraints $\lim_{n\rightarrow\infty}\frac{1}{n}\sum_{k=1}^{n}D_{k,l}\leq\bar{D}_{l}$,
collected for notational convenience in vector $\mathbf{\bar{D}}=\left[\bar{D}_{1},\ldots,\bar{D}_{L}\right]^{T}$.
As for the single sensor scenario, we are interested in finding a
distortion-optimal energy-neutral class of policies $\Upsilon^{do}\subseteq\Upsilon$,
i.e., a subset of all possible scheduling policies $\Upsilon$ that
contains at least one $\mathbf{\bar{D}}$-feasible policy.

In the following we state a necessary condition for the existence
of a $\mathbf{\bar{D}}$-feasible policy $\upsilon$.
\begin{prop}
\label{pro:scheduling_nec}For a set of distortion constraints $\mathbf{\bar{D}}$,
a necessary condition for the existence of a \textup{$\mathbf{\bar{D}}$}-feasible
policy $\upsilon$ is the existence of the set of parameters $D_{l}^{q_{l}}\geq0$,
$T_{s,l}^{q_{l}}\geq0$ for $q_{l}\in\mathcal{Q}_{l}$, $T_{t,l}^{h_{l}}\geq0$
for $h_{l}\in\mathcal{H}_{l}$, $0<\beta_{l}^{\mathbf{h}}<1$ , for
$\mathbf{h}\in\mathcal{H}^{L}$, and $0<\alpha_{l}<1$, such that\textup{
}\begin{equation}
\sum_{q_{l}\in\mathcal{Q}_{l}}\Pr(q_{l})f^{q_{l}}\left(D_{l}^{q_{l}},T_{s,l}^{q_{l}}\right)<\sum_{h_{l}\in\mathcal{H}_{l}}\left(\sum_{\mathbf{h}\in\mathcal{H}^{L}:\mathbf{h}\left(l\right)=h_{l}}\Pr(\mathbf{h})\beta_{l}^{\mathbf{h}}\right)g^{h_{l}}\left(T_{t,l}^{h_{l}}\right),\label{eq:sched_nec_cond}\end{equation}
\begin{equation}
\sum_{l=1}^{L}\beta_{l}^{\mathbf{h}}=1\mbox{ for all }\mathbf{h},\;\;\;\;\;\;\;\;\;\sum_{q_{l}\in\mathcal{Q}_{l}}\Pr(q_{l})D_{l}^{q_{l}}\leq\bar{D}_{l}\label{eq:sched_dist_cond}\end{equation}
\hspace{-1cm}\begin{equation}
\sum_{q_{l}\in\mathcal{Q}_{l}}\Pr(q_{l})T_{s,l}^{q_{l}}\leq\left(1-\alpha_{l}\right)\mathbb{E}\left[E_{l,k}\right],\;\mbox{ and }\;\sum_{h_{l}\in\mathcal{H}_{l}}\left(\sum_{\mathbf{h}\in\mathcal{H}^{L}:\mathbf{h}\left(l\right)=h_{l}}\Pr(\mathbf{h})\beta_{l}^{\mathbf{h}}\right)T_{t,l}^{h_{l}}\leq\alpha_{l}\mathbb{E}\left[E_{l,k}\right].\label{eq:sched_channel_cond}\end{equation}
\end{prop}
\begin{rem}
The interpretation of Proposition \ref{pro:scheduling_nec} is similar
to the one of Proposition \ref{pro:necessary-conditionTDO} given
in Remark \ref{rem:TDO}. The additional parameters $\beta_{l}^{\mathbf{h}}$
can be interpreted as the fraction of the subset of time slots with
joint channel state equal to $\mathbf{h}$ for which the sensor $l$
is scheduled.\end{rem}
\begin{IEEEproof}
A necessary condition for the stability of the queue of the $l^{th}$
sensor is $\mathbb{E}_{\upsilon}\left[f\left(D_{l,k},T_{s,l,k},Q_{l,k}\right)\right]<\mathbb{E}_{\upsilon}\left[g\left(H_{l,k},T_{t,l,k}\right)\right]$
\cite{BOR76}. Since we have $T_{t,l,k}=0$ for $\tau_{k}\neq l$,
on the right of the inequality we obtain $\mathbb{E}_{\upsilon}\left[g\left(H_{l,k},T_{t,l,k}\right)\right]=\sum_{h\in\mathcal{H}_{l}}\sum_{\mathbf{h}\in\mathcal{H}^{L}:\mathbf{h}\left(l\right)=h}\Pr_{\upsilon}\left(\tau_{k}=l,\mathbf{H}_{k}=\mathbf{h}\right)\times$\linebreak$\times\mathbb{E}_{\upsilon}\left[g\left(H_{l,k},T_{t,l,k}\right)|\tau_{k}=l,H_{l,k}=h\right]$.
Then, using Jensen inequality at both sides, the stability condition
becomes\begin{eqnarray}
\sum_{q_{l}\in\mathcal{Q}_{l}}\Pr(q_{l})f^{q_{l}}\left(\mathbb{E}_{\upsilon}\left[D_{l,k}\mid Q_{l,k}=q_{l}\right],\mathbb{E}_{\upsilon}\left[T_{s,l,k}\mid Q_{l,k}=q_{l}\right]\right)\leq\mathbb{E}_{\upsilon}\left[f\left(D_{l,k},T_{s,l,k},Q_{l,k}\right)\right]<\nonumber \\
<\sum_{h_{l}\in\mathcal{H}_{l}}\sum_{\mathbf{h}\in\mathcal{H}^{L}:\mathbf{h}\left(l\right)=h_{l}}\Pr_{\;\;\;\;\upsilon}\left(\tau_{k}=l,\mathbf{H}_{k}=\mathbf{h}\right)\mathbb{E}_{\upsilon}\left[g\left(H_{l,k},T_{t,l,k}\right)|\tau_{k}=l,H_{l,k}=h_{l}\right] & \leq\label{eq:stab_cond_proof_sched}\\
\leq\sum_{h_{l}\in\mathcal{H}_{l}}\sum_{\mathbf{h}\in\mathcal{H}^{L}:\mathbf{h}\left(l\right)=h_{l}}\Pr(\mathbf{h})\Pr_{\;\;\;\;\upsilon}\left(\tau_{k}=l|\mathbf{H}_{k}=\mathbf{h}\right)g^{h_{l}}\left(\mathbb{E}_{\upsilon}\left[T_{t,l,k}|\tau_{k}=l,H_{l,k}=h_{l}\right]\right).\nonumber \end{eqnarray}
Finally, condition (\ref{eq:sched_nec_cond}) is obtained for $D_{l}^{q_{l}}=\mathbb{E}_{\upsilon}\left[D_{l,k}\mid Q_{l,k}=q_{l}\right]$,
$T_{s,l}^{q_{l}}=\mathbb{E}_{\upsilon}\left[T_{s,l,k}\mid Q_{l,k}=q_{l}\right]$,\linebreak$\beta_{l}^{\mathbf{h}}=\Pr_{\upsilon}\left(\tau_{k}=l|\mathbf{H}_{k}=\mathbf{h}\right)$,
and $T_{t,l}^{h_{l}}=\mathbb{E}_{\upsilon}\left[T_{t,l,k}|\tau_{k}=l,H_{l,k}=h_{l}\right]$.
Note that (\ref{eq:sched_dist_cond})-left follows immediately from
this definition. As for conditions (\ref{eq:sched_channel_cond}),
recall that, from (\ref{eq:coda1}), we must have $\frac{1}{K}\sum_{k=1}^{K}\left(T_{s,l,k}+T_{t,l,k}\right)\leq\frac{1}{K}\sum_{k=1}^{K}E_{l,k}+\frac{\tilde{E}_{l,0}}{K},\;\mbox{for }K\geq1$,
and the initial state of the energy buffer $\tilde{E}_{l,0}$. Thus,
for a stationary ergodic policy $\upsilon$, we get \begin{equation}
\mathbb{E}_{\upsilon}\left[T_{s,l,k}\right]+\sum_{h_{l}\in\mathcal{H}_{l}}\sum_{\mathbf{h}\in\mathcal{H}^{L}:\mathbf{h}\left(l\right)=h_{l}}\Pr(\mathbf{h})\Pr_{\;\;\;\;\upsilon}\left(\tau_{k}=l|\mathbf{H}_{k}=\mathbf{h}\right)\mathbb{E}_{\upsilon}\left[T_{t,l,k}|\tau_{k}=l,H_{l,k}=h_{l}\right]\leq\mathbb{E}\left[E_{l,k}\right],\label{eq:energy_cond_sched}\end{equation}
where $\frac{1}{K}\sum_{k=1}^{K}T_{s,l,k}\rightarrow\mathbb{E}_{\upsilon}\left[T_{s,l,k}\right]$,
$\frac{1}{K}\sum_{k=1}^{K}E_{l,k}+\frac{\tilde{E}_{l,0}}{K}\rightarrow\mathbb{E}\left[E_{l,k}\right]$,and
$\frac{1}{K}\sum_{k=1}^{K}T_{t,l,k}\rightarrow\mathbb{E}_{\upsilon}\left[T_{t,l,k}\right]=$\linebreak$=\sum_{h_{l}\in\mathcal{H}_{l}}\sum_{\mathbf{h}\in\mathcal{H}^{L}:\mathbf{h}\left(l\right)=h_{l}}\Pr(\mathbf{h})\Pr_{\upsilon}\left(\tau_{k}=l|\mathbf{H}_{k}=\mathbf{h}\right)\mathbb{E}_{\upsilon}\left[T_{t,l,k}|\tau_{k}=l,H_{l,k}=h_{l}\right]$.
Given the above definition of $\beta_{l}^{\mathbf{h}}$, $T_{s,l}^{q_{l}}$
and $T_{t,l}^{h_{l}}$, inequality (\ref{eq:energy_cond_sched}) becomes
$\sum_{q_{l}\in\mathcal{Q}_{l}}\Pr(q_{l})T_{s,l}^{q_{l}}+$\linebreak$+\sum_{h_{l}\in\mathcal{H}_{l}}\sum_{\mathbf{h}\in\mathcal{H}^{L}:\mathbf{h}\left(l\right)=h_{l}}\Pr(\mathbf{h})\beta_{l}^{\mathbf{h}}T_{t,l}^{h_{l}}\leq\mathbb{E}\left[E_{l,k}\right]$,
proving (\ref{eq:sched_channel_cond}), where $\alpha_{l}=\mathbb{E}_{\upsilon}\left[T_{t,l,k}\right]/\mathbb{E}\left[E_{l,k}\right]$.
To conclude, condition (\ref{eq:sched_dist_cond})-right follows from
the distortion constraints similar to Proposition \ref{pro:necessary-conditionTDO}.
\end{IEEEproof}
In order to define a distortion-optimal energy-neutral class of policies,
Proposition \ref{pro:scheduling_nec} suggests to consider a class
of scheduling policies $\Upsilon^{do}$ in which scheduling is done
opportunistically based on the channel states $\mathbf{h}$ of all
sensors according to a probability distribution $\beta_{l}^{\mathbf{h}}$:
if the channels are equal to $\mathbf{h}$, then sensor $l$ is selected
with probability $\beta_{l}^{\mathbf{h}}$. Notice that scheduling
is independent of the observation qualities in a given slot. Moreover,
energy and distortion allocations at each sensors are similar to the
policies $\boldsymbol{\pi}_{k}$ in the class $\Pi^{do}$ (\ref{eq:policy_TO}),
and, thus, in particular perform separate resource allocation over
source and channel encoders. It can be shown that, similar to Proposition
\ref{pro:Sufficient-conditions-TDO}, the so defined class of policies
$\Upsilon^{do}$ is distortion-optimal energy-neutral in $\Upsilon$.

\subsection{Numerical Results \label{sub:Numerical-Results_sched}}

\begin{figure}[t]
\centering{\includegraphics[width=10cm]{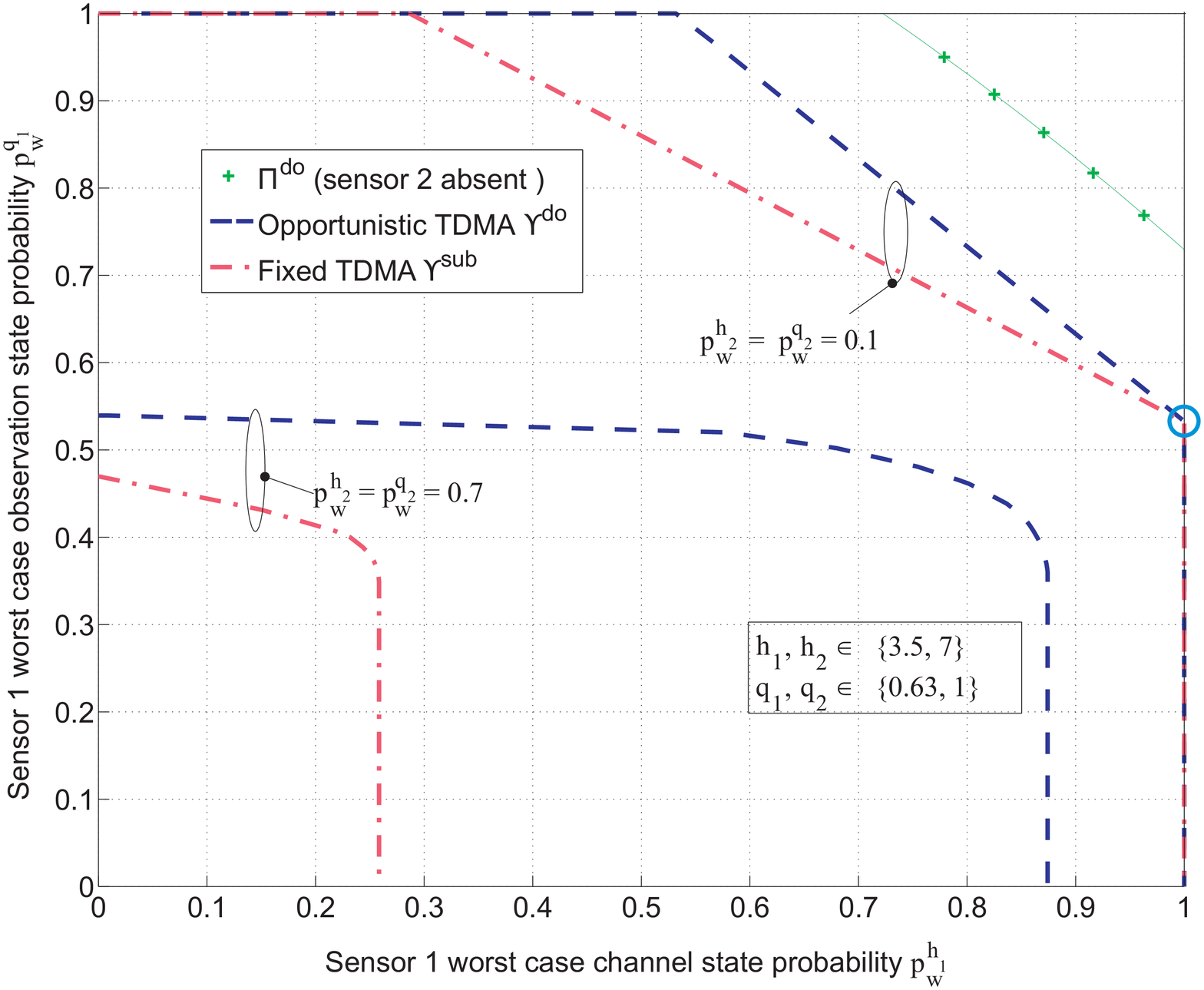}}\caption{Achievable regions (regions below the curves) of the scheduling policies
$\Upsilon^{do}$ (dashed lines) and $\Upsilon^{sub}$ (dot-dashed
lines). The solid line corresponds to the optimal policies when the
second sensor $l=2$ is not present ($E_{l,k}\sim\mathcal{U}\left(0,2\right)$,
with mean 1 Joule/channel use; $\bar{D_{l}}=0.8$; compression model
(\ref{eq:f_1}), with, for both sensors, $T_{s}^{max}=1$ Joule/source
sample, $\zeta=1$, $\eta=1.5$, bandwidth ratio $b=1$, maximum distortion
$d_{max}\mbox{=}1$).}

\label{Flo:sched_reg}%
\end{figure}

In this section we assess numerically the performance of the distortion-optimal
energy-neutral class of scheduling policies $\Upsilon^{do}$. For
comparison purposes, we introduce the suboptimal class of policies
$\Upsilon^{sub}$, that schedules each sensor according to a fixed
probability $\beta_{l}=\Pr\left(\tau_{k}=l\right)$, independently
of the current channel conditions. We consider $L=2$ sensors, which
are modeled as in Fig. \ref{Flo:stab_two_states}-b (see Sec. \ref{sub:Stability-regions:-numerical}).
For sensor $2$, the probabilities $(p_{w}^{q_{2}},p_{w}^{h_{2}})$
of the worst observation and channel states are fixed, whereas, for
sensor $1$, $(p_{w}^{q_{1}},p_{w}^{h_{1}})$ are varied. Fig. \ref{Flo:sched_reg}
shows the corresponding achievability region, defined, as in Sec.
\ref{sub:Stability-regions:-numerical}, as the set of $(p_{w}^{q_{1}},p_{w}^{h_{1}})$
for which the given policy is able to stabilize the data queues and
guarantee the given average distortions. For further comparison, Fig.
\ref{Flo:sched_reg} also shows the outer bound to the achievability
region given by the case where only sensor $1$ is present. The numerical
results confirm that the achievability region of the optimal class
of strategies $\Upsilon^{do}$ (dashed lines) is larger than that
of the suboptimal class $\Upsilon^{sub}$ (dot-dashed lines). Moreover,
note that the achievability regions shrink if the worst case states
probabilities $(p_{w}^{q_{2}},p_{w}^{h_{2}})$ of the second sensor
get larger, since sensor $2$ requires more transmission resources
to compensate for both the worst observation and channel conditions.
It is further interesting to observe that, for $p_{w}^{h_{2}}=0.1$
and $p_{w}^{h_{1}}=1$, the achievability regions of $\Upsilon^{do}$
and $\Upsilon^{sub}$, in terms of $p_{w}^{q_{1}}$, are practically
the same (circle marker). This is due to the fact that the variations
of channels $\mathbf{H}_{k}$ are not large enough to enables gains
by adapting parameters $\beta_{l}^{\mathbf{h}}$.

\section{Conclusions}

We studied energy management for a system consisting of a single sensor
whose task is that of reporting the measure of a phenomenon to a receiver.
The main problem is that of allocating energy between the source and
the channel encoders based on the current amount of available energy,
state of the data queue, quality of the measurement and of the wireless
channel. We first look for a distortion-optimal energy-neutral subset
of all policies, that contains at least one policy able to stabilize
the data queue and to satisfy a maximum average distortion constraint.
We found that optimal policies according to this criterion operate
a separate energy allocation of source and channel encoder. Instead,
we showed that a joint energy management over source and channel encoder
is required to achieve the desired trade-off between delay and distortion.
Finally, we considered a system with multiple sensors and obtained
TDMA scheduling policies that guarantee the stability of all data
queues, whenever the distortion constraints are feasible. Overall,
our results, which also include further comparisons with a number
of suboptimal policies, shed light on the challenges and design issues
that characterize modern cyber-physical systems.

\section*{Acknowledgments}

The authors would like to thank Prof. Petar Popovski of Aalborg University
for suggesting the analysis of the analog transmission scheme defined
in Sec. \ref{sub:Analog-transmission}.

\bibliographystyle{ieeetr}
\bibliography{BIB_WiOpt}

\pagebreak{}\newpage{}\clearpage{}

\pagebreak{}\newpage{}\clearpage{}

\pagebreak{}\newpage{}\clearpage{}

\pagebreak{}\newpage{}\clearpage{}

\pagebreak{}\newpage{}\clearpage{}

\end{document}